\documentclass[aps, prd,  showpacs, preprintnumbers, twocolumn, nofootinbib, superscriptaddress]{revtex4-1}

\pdfoutput=1 
\usepackage{amssymb,amsmath,hyperref}
\usepackage{color}
\usepackage{graphicx}
\usepackage{bm}
\usepackage{amsmath,amssymb}
\usepackage{graphicx}
\usepackage{bm}
\usepackage{comment}
\usepackage{color}
\usepackage{subfigure}
\usepackage{array}
\usepackage{multirow} 

\newcommand{\gsim}{\raisebox{-0.7ex}{$\stackrel{\textstyle >}{\sim}$ }}
\newcommand{\lsim}{\raisebox{-0.7ex}{$\stackrel{\textstyle <}{\sim}$ }}
\newcommand{\nn}{\nonumber}

\usepackage[T1]{fontenc}
\usepackage[latin9]{inputenc}
\usepackage{graphicx}
\usepackage{esint}
\usepackage{hyperref}
\usepackage{atbegshi}
\usepackage{lipsum}

\begin{document}

 \preprint{\vbox{
\hbox{JLAB-THY-14-1833} 
}}
\pacs{}

\title{Two-particle multichannel systems in a finite volume with arbitrary spin }
\author{Ra\'ul A. Brice\~no}
\email{rbriceno@jlab.org}

\affiliation{Jefferson Laboratory, 12000 Jefferson Avenue, Newport
  News, VA 23606, USA}
  
  \date{\today}

\begin{abstract} 
 The quantization condition for two-particle systems with arbitrary number of  two-body open coupled channels, spin, momentum, and masses in a finite volume with either periodic or twisted boundary conditions is presented. Although emphasis is placed in cubic volumes, the result holds for asymmetric volumes. The result is relativistic, holds for all momenta below the three- and four-particle thresholds, and is exact up to exponential volume corrections that are governed by $L/r$, where $L$ is the spatial extent of the volume and $r$ is the range of the interactions between the particles. For hadronic systems the range of the interaction is set by the inverse of the pion mass, $m_\pi$, and as a result the formalism presented is suitable for $m_\pi L\gg1$. The condition presented is in agreement with all previous studies of two-body systems in a finite volume. Implications of the formalism for the studies of multichannel baryon-baryon systems are discussed. 
 \end{abstract}
\maketitle
\section{Introduction  \label{sec:Intro} }

There is a wealth of experimental investigation of low-energy scattering processes involving two hadrons. Yet to this day, comparison of the available experimental data with the underlying fundamental theory of the strong interaction, quantum chromodynamics (QCD), has been limited. This is due to the fact that at moderately low energies QCD is non-perturbative. Currently the only reliable approach for studying QCD at low energies is lattice QCD (LQCD). LQCD calculations are necessarily performed in a finite Euclidean spacetime. Therefore, it is necessary to construct a formalism that connects the finite Euclidean spacetime volume observables determined via LQCD to the  Minkowski-spacetime infinite-volume quantities of interest. The most easily determined quantities via LQCD are the low-energy spectra. For sufficiently large volumes satisfying $m_\pi L\gsim 4$, where $L$ is the spatial extent of the finite volume and $m_\pi$ is the pion mass, finite volume effects have minimal impact in the determination of the low-lying single-particle spectrum of QCD~\cite{Luscher:1985dn}. 

Although it has been previously pointed out that the Euclidean nature of the calculations imposes challenges on the determination of few-body scattering quantities for arbitrary momenta in the infinite volume limit \cite{Maiani:1990ca}, the fact that LQCD calculations are performed in a \textit{finite volume} (FV) allows for the extraction of scattering parameters from the spectrum through the \textit{L\"uscher} method \cite{Luscher:1986pf, Luscher:1990ux, Rummukainen:1995vs,  Beane:2003yx, Beane:2003da, Li:2003jn, Detmold:2004qn, Feng:2004ua, Christ:2005gi, Kim:2005gf, Bernard:2008ax, Bour:2011ef, Davoudi:2011md, Leskovec:2012gb, Gockeler:2012yj, Ishizuka:2009bx, Briceno:2013lba}. This method, which has been widely used to extract scattering phase shifts and binding energies of two-hadron systems from LQCD (see for example Refs. \cite{He:2005ey, Li:2007ey, Durr:2008zz, Beane:2010hg, Beane:2011xf, Beane:2012vq, Beane:2012ey,Yamazaki:2012hi, Beane:2011iw, Beane:2013br, Beane:2011sc,  Pelissier:2011ib, Aoki:2007rd, Lang:2011mn, Pelissier:2012pi, Ozaki:2012ce, Buchoff:2012ja, Dudek:2012xn, Dudek:2012gj, Mohler:2013rwa, Lang:2014tia}), has been generalized to multi-coupled channel two-body systems with total spin $S\leq1/2$~\cite{Liu:2005kr, Hansen:2012tf, Briceno:2012yi,  Li:2012bi, Guo:2012hv, Bernard:2010fp}. There has also been some progress in generalizing this formalism onto three-particle systems~\cite{Roca:2012rx,  Polejaeva:2012ut,Briceno:2012rv, Hansen:2013dla}.

Although LQCD calculations are commonly performed with the periodic boundary conditions (PBCs) imposed upon the quark fields in the spatial extents, PBCs are a subset of a more general class of boundary conditions known as twisted boundary condition (TBCs)~\cite{PhysRevLett.7.46, Bedaque:2004kc}. TBCs require that fields are proportional to their images up to an overall phase,  
${\psi}(\mathbf{x}+\mathbf{n}{{L}})=e^{i{\bm{\theta}} \cdot \mathbf{n}}{\psi}(\mathbf{x}),$ 
where $0\leq\theta_i<2\pi$ is the twist angle in the $ith$ Cartesian direction. As a result the free finite volume momenta satisfy $\mathbf{p}=\frac{2\pi}{L}\mathbf{n}+\frac{\bm{\theta}}{L}$,
where $\mathbf{n}$ is an integer triplet. PBCs are recovered when the twist angle, $\bm{\theta}$, is set to zero. It is evident that, at least in the one-body sector, by dialing the twist one can in principle access a continuous set of momenta. This is advantageous when performing calculations in a finite volume where spectra are necessarily discretized, and has been explored extensively in the one-body sector~\cite{deDivitiis:2004rf, Sachrajda:2004mi, Tiburzi:2005hg,Jiang:2006gna,Boyle:2007wg,Simula:2007fa,Boyle:2008yd,Aoki:2008gv, Brandt:2013mb} as well as the two-body sector~\cite{Bedaque:2004kc, Bedaque:2004ax, Bernard:2010fp, Doring:2011vk, Ozaki:2012ce, Briceno:2013hya, Agadjanov:2013wqa}.

Here we remove all previous restrictions made in the literature and present the most general, model-independent relativistically covariant framework for determining the finite volume (FV)  spectrum for two-particle multichannel systems with arbitrary spin, momenta and twist. Although this formalism is developed with LQCD calculations in mind, the result gives a mapping between the finite volume spectrum and the infinite volume scattering amplitude of the system and is independent of the details of the theory at hand. It is suitable for studying hadronic physics as well as atomic physics in a finite volume~(see Refs.~\cite{Drut:2012md, Endres:2012cw} for examples of the L\"uscher method applied on atomic systems). 

Section~\ref{sec:helicityLS} reviews basic tools for constructing two-particle states with arbitrary spin. In particular, we review details regarding the $|lS,Jm_J\rangle$ and helicity bases\footnote{All throughout this work, $J$ will denote the total angular momentum, $l$ is the orbital angular momentum, $S$ will be the two-particle spin, and $m_J$ is the azimuthal component of $J$. Capital $``L"$ will solely refer to the spatial extent of the finite volume. The $|lS,Jm_J\rangle$ basis will be referred to as as the \emph{lS} basis.}. Section~\ref{sec:SpinFVPBCs} presents the generalization of the L\"uscher formalism for two-particle systems with arbitrary quantum numbers, masses and momenta in a periodic finite volume and we pay close attention to the evaluation of finite volume s-channel loops. Section~\ref{sec:SpinFVTBCs} discusses how this result is generalized for systems with arbitrary twist and asymmetry volumes. The result presented, Eq.~\ref{eq:QC}, is in agreement with all previous studies of two-body systems in a finite volume~\cite{Luscher:1986pf, Luscher:1990ux, Rummukainen:1995vs, Feng:2004ua, Christ:2005gi, Kim:2005gf, Bernard:2008ax, Bour:2011ef, Davoudi:2011md, Leskovec:2012gb,  Gockeler:2012yj, He:2005ey, Liu:2005kr, Hansen:2012tf, Briceno:2012yi,  Li:2012bi, Guo:2012hv, Bernard:2010fp, Ishizuka:2009bx, Briceno:2013lba, Bernard:2010fp, Doring:2011vk, Ozaki:2012ce, Briceno:2013hya, Agadjanov:2013wqa}. Section~\ref{sec:baryon2} reviews the implication of this formalism for baryon-baryon system, and we discuss its impact on future studies of hyperon-nucleon  and hyperon-hyperon scattering parameters from LQCD. Precise determination of such interactions will impact our understanding of the composition of dense nuclear matter. Furthermore, although there has been a great deal of activity at elucidating the poor signal/noise problem that is inherit of performing LQCD calculation with finite baryon density and/or chemical potential~\cite{Lepage89, MJSsign, Grabowska:2012ik,Kaplan:2013dca}, nucleon-nucleon LQCD calculations remain to be more computationally costly than those with higher strange content. Consequently, it is expected that LQCD calculations will have a bigger immediate impact in disentangling pertinent information of nuclear systems with non-zero strangeness.


\section{Construction of two-particle states with arbitrary spin \label{sec:helicityLS} }
 
In order to understand the claim that the results presented in Secs.~\ref{sec:SpinFVPBCs} \& \ref{sec:SpinFVTBCs} are covariant and applicable for systems with arbitrary spin it is important to first review the basics of the construction of two-particle states in the $lS$ basis as well as the helicity basis. In order to construct single particle states with arbitrary helicity $\lambda$ and momentum $\textbf{p}=p~(\sin\theta\cos\phi,\sin\theta\sin\phi,\cos\theta)$, it is convenient to first define a state with zero total momentum, definite spin (s) and azimuthal component of spin ($\lambda$), $|\textbf{0},s,\lambda\rangle$. By acting on this state with a boost along the z-axis, $\hat{L}_z(p)$, followed by a rotation, $\hat{R}_{\phi,\theta,-\phi}$\footnote{In general a rotation can be defined by a unitary operator with three angles as arguments $\hat{R}_{\alpha,\beta,\gamma}=e^{-i\alpha \hat{J}_z}e^{-i\beta \hat{J}_y}e^{-i\gamma \hat{J}_z}$, where $\hat{J}_k$ is the angular momentum operator in the $kth$ cartesian axis. To define a three-dimensional vector only  two angles are needed. As a result there is some freedom when defining the rotation operator. In this work the operators is chosen to be $\hat{R}_{\phi,\theta,-\phi}$, such that when $\theta=0$ and it acts on a state with angular momentum quantized along the z-axis the overall phase vanishes.}, that takes the momentum from the z-axis to the desired direction of the momentum, one obtains the desired state with definitive helicity \cite{Jacob:1959at, :/content/aip/journal/jmp/4/4/10.1063/1.1703981, McKerrell64,  Chung:2007nn}
\begin{eqnarray}
|\textbf{p},s \lambda\rangle&=&\hat{R}_{\phi,\theta,-\phi}~\hat{L}_z(p)|\textbf{0},s \lambda\rangle.
\end{eqnarray} 
Two-particles states can be built out of direct product of these, 
\begin{eqnarray}
|\textbf{p}_1,s_1\lambda_1;\textbf{p}_2,s_2\lambda_2\rangle&=&|\textbf{p}_1,s_1\lambda_1\rangle\otimes|\textbf{p}_2,s_2\lambda_2\rangle.\nn\end{eqnarray} 
When restricting oneself to the center of mass (c.m.) frame this simplifies to\footnote{In general, this describes the component of the wavefunction that only depends on the relative coordinates.}
\begin{eqnarray}
|\textbf{q}^*,\lambda_1\lambda_2\rangle&=&|\textbf{q}^*,s_1\lambda_1\rangle\otimes|-\textbf{q}^*,s_2 \lambda_2\rangle
\nn\\
&=&\hat{R}_{\phi,\theta,-\phi}~\hat{K}_z({q}^*)|\textbf{0},s_1\lambda_1\rangle\nn\\
&&\otimes
\hat{R}_{\phi,\theta,-\phi}~\hat{K}_{-z}({q}^*)|\textbf{0},s_2\lambda_2\rangle,
\end{eqnarray} 
where $\textbf{q}^*$ is the relative momenta between the two-particles in the c.m. frame\footnote{Center of mass frame coordinates and functions will be given a superscript $``*"$ to distinguish them from the lattice frame coordinates.}. Note that the explicit $s_1$ and $s_2$ labels have been suppressed. From these states one can readily construct states with definite total angular momentum $(J,m_J)$~\cite{Jacob:1959at, :/content/aip/journal/jmp/4/4/10.1063/1.1703981, McKerrell64,  Chung:2007nn}
\begin{eqnarray}
|Jm_J,\lambda_1\lambda_2\rangle&=&\nn\\
&&\hspace{-1cm}{N_J}\int d\Omega~\mathcal{D}^{J*}_{m_J,\lambda}(\phi,\theta,-\phi)|\textbf{q}^*,\lambda_1\lambda_2\rangle,
\label{JMl1l2}
\end{eqnarray} 
where $d\Omega=\sin\theta~d\theta~d\phi$, $\lambda=\lambda_1-\lambda_2$, $\mathcal{D}^{J*}_{m_J,\lambda}(\phi,\theta,-\phi)$ is the complex conjugate of the Wigner-$\mathcal{D}$ matrix  defined as $\mathcal{D}^{J}_{m_J,\lambda}(\phi,\theta,-\phi)=\langle Jm_J|\hat{R}_{\phi,\theta,-\phi}|J\lambda\rangle$, and $N_J=\sqrt{\frac{2J+1}{4\pi}}$ as to ensure that the states are properly normalized. It is straightforward to show that in fact these states transform as states with definite angular momentum under rotations
\begin{eqnarray}
 \hat{R}_{\alpha,\beta,\gamma}|Jm_J,\lambda_1\lambda_2\rangle&=&\nn\\
&&\hspace{-1cm}\sum_{m'}\mathcal{D}^{J}_{m_{J'},m_J}({\alpha,\beta,\gamma})|Jm_{J'},\lambda_1\lambda_2\rangle\nn.
\end{eqnarray}

Although it is customary to use the helicity basis when considering relativistic systems, one can always perform calculation in the $lS$ basis, $|lS,Jm_J\rangle$. In order to properly define these states, one may first define the angular momentum operator $\hat{\bm J}$ as a sum of the spin and orbital angular momentum operators, $\hat{\bm J}=\hat{\bm S}+\hat{\bm{l}}$~\cite{:/content/aip/journal/jmp/4/4/10.1063/1.1703981, McKerrell64}. It has been shown that the spin operator can be written as a combination of all the generators of the Poincar\'e group, i.e., $\hat {\bm J}$ (rotations), $\hat {\bm K}$ (boosts) and $( \hat P^0,{\hat {\bm P}})$ (translations)~\cite{:/content/aip/journal/jmp/4/4/10.1063/1.1703981, McKerrell64}
\begin{eqnarray}
\hat {\bm{S}}=\frac{1}{M}\left(\hat {P}^0 \hat {\bm {J}}-\hat {\bm{P}}\times\hat {\bm{K}} -\frac{1}{\hat {P}^0+M}\hat {\bm {P}}~(\hat {\bm{P}}\cdot \hat {\bm {J}})\right),~~
\end{eqnarray}
where $M$ is the mass of the particle of interest. The orbital angular momentum operator can then defined as $\hat{\bm l}=\hat{\bm J}-\hat{\bm S}$. Having defined these two operators, one can perform two distinct classes rotations, the first where $\hat{\bm{S}}$ is used as the generator of the rotation $\hat{R}^S_{\alpha,\beta,\gamma}=e^{-i\alpha \hat{S}_z}e^{-i\beta \hat{S}_y}e^{-i\gamma \hat{S}_z}$ and the second for $\hat{\bm{l}}$ $\hat{R}^l_{\alpha,\beta,\gamma}=e^{-i\alpha \hat{l}_z}e^{-i\beta \hat{l}_y}e^{-i\gamma \hat{l}_z}$. As one would expect, these operators can be interpreted as acting on the spin and spatial components of the single-particle states respectively~\cite{McKerrell64}, 
\begin{eqnarray}
\hat{R}^S_{\phi,\theta,-\phi}|\textbf{p},sm_{s}\rangle&=&\sum_{m_{s'}}
\mathcal{D}^{s}_{m_{s'},m_{s}}(\phi,\theta,-\phi)|\textbf{p},sm_{s'}\rangle,\nn\\
\hat{R}^l_{\phi,\theta,-\phi}|\textbf{p},sm_{s}\rangle&=&|\textbf{p}'=\hat{R}~\textbf{p},sm_{s}\rangle,
\label{eq:LSrot}
\end{eqnarray}
where $\hat{R}$ is the three-dimensional representation of the rotation acting on $\textbf{p}$. Effectively one can conclude that the $\hat{R}^S_{\phi,\theta,-\phi}$ operator acts on a relativistic state as if it were at rest, while the $\hat{R}^l_{\phi,\theta,-\phi}$ operator acts on it as if it had zero spin. 

For two-particle systems this can be generalized by defining the two body spin and orbital angular momentum operators
\begin{eqnarray}
\hat{\bm{S}}=\hat{\bm{S}}_1+\hat{\bm{S}}_2,\hspace{.5 cm}
\hat{\bm{l}}=\hat{\bm{l}}_1+\hat{\bm{l}}_2,
\end{eqnarray}
where $\hat{\bm{S}}_1(\hat{\bm{S}}_2)$ and $\hat{\bm{l}}_1(\hat{\bm{l}}_2)$ are, respectively, the spin and orbital angular momentum operators that act on the $``1"(``2")$ particle state. By restricting oneself to the c.m. frame, the two-particle state with total spin $S$ can be defined in terms of single particle states in the standard way, 
\begin{eqnarray}
|\textbf{q}^*,Sm_S\rangle&=&\sum_{m_{s_1},m_{s_2}}|\textbf{q}^*,s_1m_{s_1}\rangle\otimes|-\textbf{q}^*,s_2m_{s_2}\rangle\nn\\
&&\hspace{.5cm}\times \langle s_1m_{s_1},s_2m_{s_2} |s_1s_2,Sm_S\rangle, \nn
\end{eqnarray} 
where $\langle s_1m_{s_1},s_2m_{s_2} |s_1s_2,Sm_S\rangle$ is the Clebsch-Gordan coefficient. Similarly to the one-particle system, one can show that under $R^S$ and $R^l$ these states transform as 
\begin{eqnarray}
\hat{R}^S_{\phi,\theta,-\phi}|\textbf{q}^*,Sm_{S}\rangle&=&\sum_{m_{S'}}
\mathcal{D}^{S}_{m_{S'},m_{S}}(\phi,\theta,-\phi)|\textbf{q}^*,Sm_{S'}\rangle,\nn\\
\hat{R}^l_{\phi,\theta,-\phi}|\textbf{q}^*,sm_{S}\rangle&=&|{\textbf{q}^*}'=\hat{R}~\textbf{q}^*,Sm_{S}\rangle.
\label{eq:LSrot}
\end{eqnarray}
States with definite orbital angular momentum can be constructed by integration over all angles of the relative momentum with the appropriate spherical harmonic
\begin{eqnarray}
|lm_l,Sm_S\rangle&=&
\int d\Omega~Y_{lm_l}(\hat{\textbf{q}^*})~|\textbf{q}^*,Sm_S\rangle,\\
\Rightarrow
\hat{R}^l_{\phi,\theta,-\phi}|lm_l,Sm_S\rangle&=&\nn\\
&&\hspace{-2cm}\sum_{m_{l'}}\mathcal{D}^{l}_{m_{l'},m_{l}}(\phi,\theta,-\phi)|lm_{l'},Sm_S\rangle,
\end{eqnarray} 
where $m_l$ is the azimuthal component of the orbital angular momentum. Using Clebsch-Gordan coefficients these states can be added appropriately to give a state with total angular momentum
\begin{eqnarray}
|lS,Jm_J\rangle&=&\nn\\
&&\hspace{-1cm}
\sum_{m_l,m_S}
|lm_l,Sm_S\rangle\langle lm_l,Sm_S|lS,Jm_J\rangle.
\label{JMlsbasis}
\end{eqnarray} 

One can show that these states can be written as a linear combination of the states with definite helicity with an overlap factor equal to~\cite{Jacob:1959at}
\begin{eqnarray}
\langle Jm_J,\lambda_1\lambda_2 |lS,Jm_J\rangle&=&\left(\frac{2l+1}{2J+1}\right)^{1/2}
\langle l0,S\lambda|J\lambda\rangle\nn\\
&&\hspace{.5cm}
\times\langle s_1\lambda_1,s_2-\lambda_2 |s \lambda\rangle.
\label{eq:helicityLS}
\end{eqnarray}  

Since the total angular momentum is a conserved quantity, the $\bm{2}\rightarrow\bm{2}$ scattering amplitude, ${\mathcal{M}}$, is diagonal in $J$. One may choose to evaluate its matrix elements in the helicity basis using Eq.~\ref{JMl1l2}, in which case one finds
\begin{eqnarray}
&&\langle \textbf{q}_f^*,\alpha_{1}\alpha_{2}|{\mathcal{M}}|\textbf{q}_i^*,\lambda_{1}\lambda_{2}\rangle=\nn
\sum_{J,m_J}\left({N_J}\right)^2
[{\mathcal{M}}]^{Jm_J}_{\alpha_{1}\alpha_{2},\lambda_{1}\lambda_{2}}\\
&&\hspace{1cm}\times
\mathcal{D}^{J*}_{m_J,\alpha}(\phi_f,\theta_f,-\phi_f)
\mathcal{D}^{J}_{m_J,\lambda}(\phi_i,\theta_i,-\phi_i),
\end{eqnarray} 
where $[{\mathcal{M}}]^{J}_{\alpha_{1}\alpha_{2},\lambda_{1}\lambda_{2}}$ is the value of the scattering amplitude for a initial state with helicity $\lambda_1,\lambda_2$ and final helicity $\alpha_1,\alpha_2$ and that has been projected onto total angular momentum $(J,m_J)$. Alternatively, one can write the scattering amplitude in the $lS$ basis using Eq.~\ref{JMlsbasis}, 
\begin{eqnarray}
&&\langle \textbf{q}_f^*,S'm_{S'}|{\mathcal{M}}|\textbf{q}_i^*,Sm_{S}\rangle=
4\pi
\nn
\sum_{\stackrel{J,m_J,l,l'}{m_{l},m_{l'}}}
Y_{l'm_{l'}}(\hat{\textbf{q}}^*_f)
Y^*_{lm_{l}}(\hat{\textbf{q}}^*_i)
\\
&&\hspace{.0cm}\times \langle lm_lSm_s|lS,Jm_J\rangle 
\langle l'm_{l'}S'm_{s'}|l'S',Jm_J\rangle  
~[{\mathcal{M}}]^{Jm_J}_{l'S',lS}, ~~~~~~
\label{MJMlsbasis}
\end{eqnarray} 
where $[{\mathcal{M}}]^{Jm_J}_{l'S',lS}$ is the value of the scattering amplitude for an ingoing state with $(l,S)$ and outgoing $(l',S')$ and that has been projected onto total angular momentum $(J,m_J)$. The $\sqrt{4\pi}$ factor for each spherical harmonic has been introduced to simplify the subsequent expressions in Sec.~\ref{sec:loop}. 
Given that these two representations are equivalent, in the remainder of this work the $lS$ basis will be used. All that has been assumed in writing Eq.~\ref{MJMlsbasis} is that the scattering is diagonal in angular momentum. Therefore, Eq.~\ref{MJMlsbasis} holds for any quantity that is diagonal in $J$. When considering a system with N open two-body channels that can couple, one can simply upgrade the scattering amplitude to also be a matrix in the number of open channel. For such cases, the matrix elements of $\mathcal{M}$ get an additional subscript associated with the incoming ($``a"$) and outgoing ($``b"$) channel, $[\mathcal{M}]^{Jm_J}_{l'S'b,lSa}$. 

\section{Two-particle multichannel systems with spin and PBCs\label{sec:SpinFVPBCs}}
Having reviewed the basics of relativistic two-particle states with spin, one may proceed to determine the finite volume spectra of such systems. To arrive at the quantization condition for multichannel two-particle systems with arbitrary spin, masses and momenta, consider a system with total energy (momentum) equal to $E$ ($\textbf{P}$) and c.m. energy $E^*=\sqrt{E^2-P^2}$. Allow for the system to have N open channels that can mix, each composed of two-particles with masses $m_{j,1}$ and $m_{j,2}$ with $m_{j,1}\leq m_{j,2}$ and spin $s_{j,1}$ and $s_{j,2}$. Each particle can be either a fermion or a boson. The particles in the $jth$ channel can go on-shell if the c.m. energy satisfies $m_{j,1}+m_{j,2}~\lsim E^*\ll E^*_{th}$, where $E^*_{th}$ refers to the first few-particle threshold present in the theory. For instance, for a systems such as $\pi\pi-K\bar{K}$ with exact G-parity, we are restricted to energies below the four-particle threshold, while for the two-nucleon systems the energy is restricted below the pion production threshold. Furthermore, it will be assumed that no single particle states can go on-shell.  The c.m. relative momentum for the $jth$ channel satisfies~\footnote{For bound states $E^*<m_{j,1}+m_{j,2}$, which leads to the relative momentum to be imaginary $k^{*2}_j<0$. Although it may be sometimes desirable to approximate the finite volume effects associated with the determination of a bound state energy in a finite volume~\cite{Bour:2011ef, Konig:2011nz, Davoudi:2011md, Konig:2011ti}, the formalism presented here non-perturbatively describes such effects for bound states. }
\begin{eqnarray}
\label{momentumcc}
k^{*2}_j=\left(\frac{E^{*2}}{4}-\frac{(m_{j,1}^2+m_{j,2}^2)}{2}+\frac{
(m_{j,1}^2-m_{j,2}^2)^2}{4E^{*2}}\right).~~
\end{eqnarray}

\begin{figure*}[t]
\begin{center}
\subfigure[]{
\label{fig:FVprop}
\includegraphics[scale=0.35]{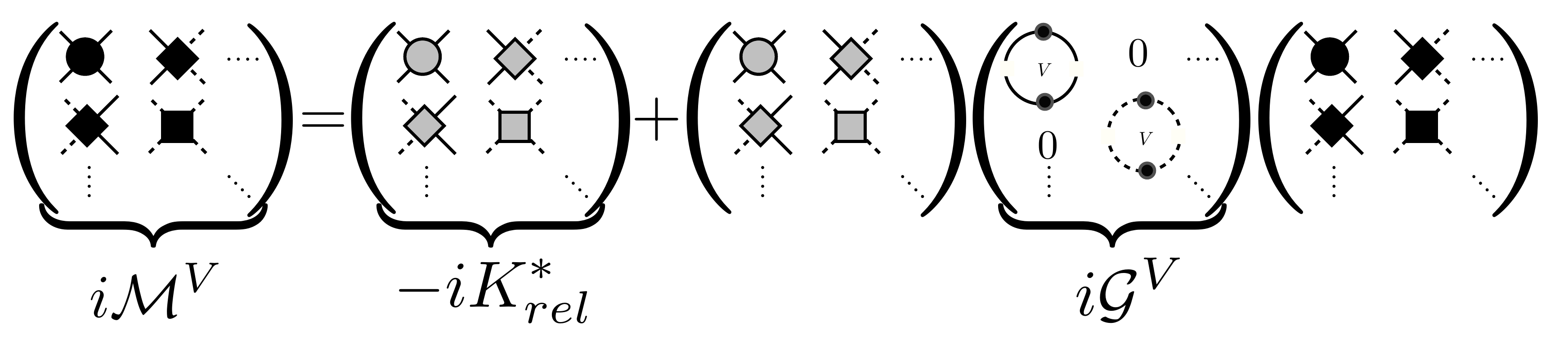}}\\
\subfigure[]{
\label{fig:kernel}
\includegraphics[scale=0.25]{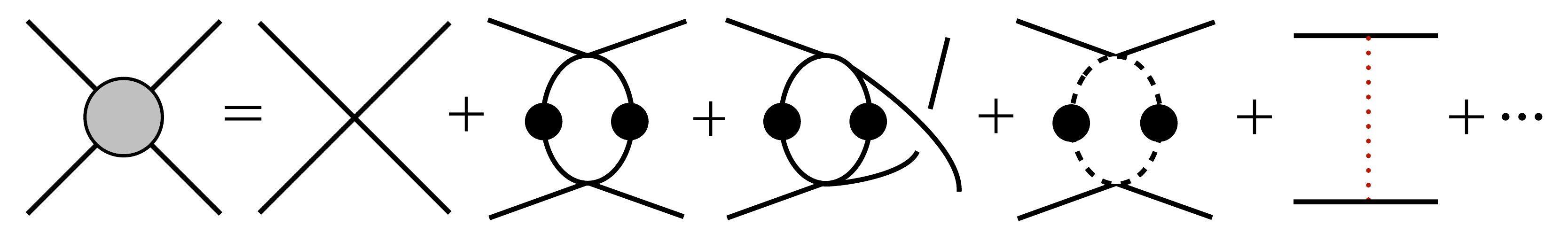}}
\subfigure[]{
\label{fig:1bodyprop}
\includegraphics[scale=0.25]{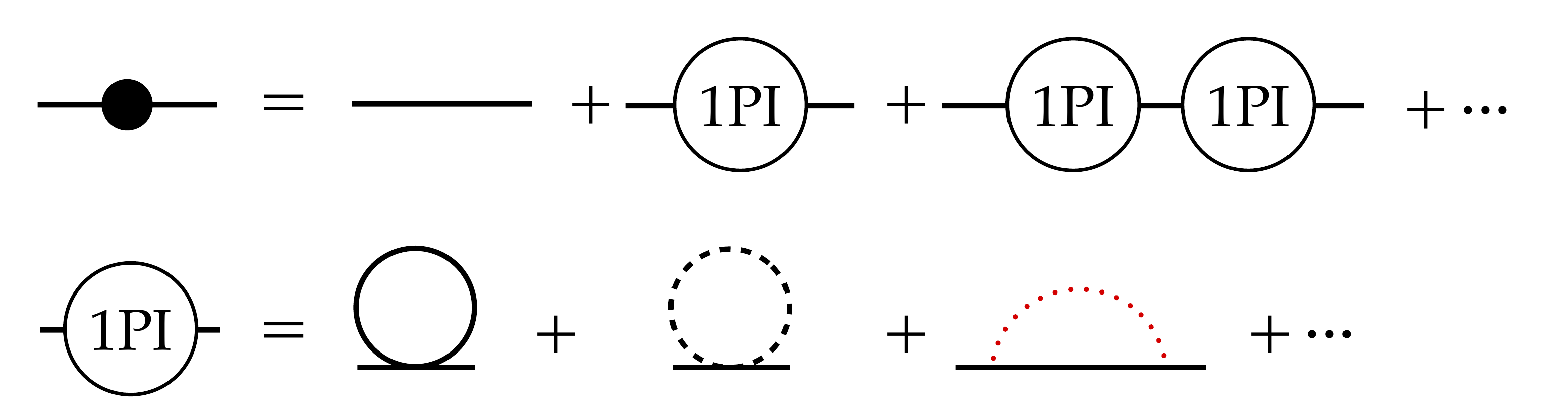}}

\caption{a) Shown is the self consistent definition of $\mathcal{M}^V$, which is defined as the sum of all $\textbf{2}\rightarrow\textbf{2}$ finite volume diagrams, Eq.~\ref{FV_prop}. The solid lines denote two-particles in the $``1"$ channel, dashed lines denote particle in the $``2"$ channel. $\mathcal{M}^V$ is written in terms of the c.m. kernel, ${K}^*_{\text{rel}}$, and the fully dressed single particle propagators. b) Shown is ${K}^*_{\text{rel}}$ for the first channel, which is the sum of all two-particle irreducible s-channel diagrams. Explicitly shown are examples of diagrams that are included in the kernel: contact interactions, t- and u-channel diagrams and possible meson exchange diagrams, if allowed by the symmetries of the system of interest. If the two initial and final states of the kernel are baryons these exchange diagrams are presented, otherwise they are not allowed by G-parity. In general, all diagrams allowed by the underlying theory where the intermediate particles cannot all simultaneously go on-shell are absorbed into the kernel. As described in the text, in this study we are restricted to energies where only two-particle states are allowed to go on-shell. c) Shown is the definition of the fully dressed one particle propagator in terms of the the one particle irreducible (1PI) diagrams.   }\label{fig:QC}
\end{center}
\end{figure*}

The derivation and details of the quantization condition for systems where the total spin of the open channels is zero has been presented in Refs.~\cite{Hansen:2012tf, Briceno:2012yi}. The remaining piece needed to arrive at the result with non-zero spin can be deduced from the $S=1/2$ and $S=1$ single channel results~\cite{Bernard:2008ax,  Gockeler:2012yj, Ishizuka:2009bx, Briceno:2013lba}. These works concluded that the distinguishing feature of the power law finite volume corrections of two-particle propagators between spinless systems and nonzero spin systems can be attributed to Clebsch-Gordan coefficients which project two-particles states with definite spin and orbital angular momentum to a state with total $J$ as shown in Eq.~\ref{JMlsbasis}. In Sec.~\ref{sec:loop} the emergence of these Clebsch-Gordan coefficients for generic spin systems will be shown. 

 Arriving at the $\textbf{2}\rightarrow\textbf{2}$ QC can be done by introduction the relativistic c.m. \emph{kernel}, ${K}^*_{\text{rel}}$, which is defined as the sum of all the two-particle irreducible s-channel diagrams. Just like the scattering amplitude, the kernel is a matrix over all the open channels and is diagonal in total angular momentum. An example of a matrix element of the kernel is illustrated in Fig.~\ref{fig:kernel}. Having defined ${K}^*_{\text{rel}}$, the infinite volume scattering amplitude satisfies a self-consistent matrix, integral equation
\begin{eqnarray}
\label{IV_prop}
i\mathcal{M}&=&-i{K}^*_{\text{rel}}
+i{K}^*_{\text{rel}}\mathcal{G}^{\infty}\mathcal{M}, 
\end{eqnarray}
where $\mathcal{G}^{\infty}$ is a diagonal matrix in the number of channels, orbital angular momentum and spin. Its $jth$ matrix element in \emph{channel space} is the infinite volume s-channel loop for the $jth$ channel. The $``ab"$ matrix element of the second term in the equation above, can be explicitly written as an integral of the form,
\begin{eqnarray}
i[{K}^*_{\text{rel}}\mathcal{G}^{\infty}\mathcal{M}]_{ab}
&\equiv &
\nn\\
&&\hspace{-1.5cm}
\int\frac{d^{4}q}{(2\pi)^4}[{K}_{\text{rel}}({p}_f,q)]_{aj}\Delta_{j}(q)[\mathcal{M}(q,{p}_i)]_{jb},~~~
\label{IV_prop2}
\end{eqnarray}
where ${K}_{\text{rel}}$ denotes the functional form of the kernel in the lattice frame, the dependence on the total four-momentum $P$ is left implicit, and summation over repeated indices is implied. $\Delta_{j}(q)$ denotes relativistic two-particle propagator. In general, this is a matrix in spin that mixes different azimuthal components of spin. In the helicity basis it diagonal and can be written as 
\begin{eqnarray}
[\Delta_{j}(q)]_{\alpha_1,\alpha_2;\lambda_1,\lambda_2}&=&\nn\\
&&\hspace{-1.5cm}\frac{z_{j,1}(P-q)~z_{j,2}(q)~\delta_{\alpha_1,\lambda_1}~\delta_{\alpha_2,\lambda_2}}{[(q-P)^2-m_{j,1}^2+i\epsilon][q^2-m_{j,2}^2+i\epsilon]} ,
\label{bareprop}
\end{eqnarray}
where $z_{j,i}$ is the residue of the $ith$ single particle, fully dressed propagator in the $jth$ channel.  

As will become evident shortly, when interested in the determination of the finite volume it will not be necessary to give an explicit expression for the infinite volume function $\mathcal{G}^{\infty}$, and all that will be necessary is to determine the difference between this object and its finite volume counterpart, $\delta\mathcal{G}^{V}=\mathcal{G}^{V}-\mathcal{G}^{\infty}$. In Sec.~\ref{sec:loop}, $\delta\mathcal{G}^{V}$ is derived for systems with periodic boundary conditions and the result is given in Eq.~\ref{eq:deltaGPBCs}, and the expression for systems with arbitrary twist and asymmetry volumes is given in Sec.~\ref{sec:SpinFVTBCs}.

 In order to define the relation between the scattering amplitude and the $S$-matrix, it is convenient to introduce a matrix that is diagonal over the N open channels $\mathbb{P}=\text{diag}(\sqrt{n_1q^{*}_1},\sqrt{n_2q^{*}_2},\ldots,\sqrt{n_Nq^{*}_N})/\sqrt{4\pi E^*}$, where $n_j$ is the symmetry factor for the $jth$ channel and is equal to $1/2$ if the two-particles are identical and 1 otherwise. The $S$-matrix is diagonal in the total angular momentum basis. For a system with total angular momentum $J$, the scattering amplitude $\mathcal{M}_{J}$ is related to the $S$-matrix for that channel via,~\cite{Hansen:2012tf}
\begin{eqnarray}
i\mathcal{M}_{J}=\mathbb{P}^{-1}~{(S_{J}-\mathbb{I})}~\mathbb{P}^{-1}.\label{Smatrix}
\end{eqnarray}
For spinless systems the orbital angular momentum is equal to the total angular momentum. For systems with nonzero spin, this will in general not be true. For instance in the spin-triplet positive parity two-nucleon channel, considered in Ref~\cite{Briceno:2013lba}, $S_1$ would be a $6\times6$ matrix that couples the ${^3}S_1$ and ${^3}D_1$ ,
\begin{eqnarray}
S_{1}&=&\left( \begin{array}{cc}
S_{1}^{S}&S_{1}^{SD}\\
S_{1}^{DS}&S_{1}^D
\label{J1Smatrix}
\end{array} \right).
\end{eqnarray}
Each of the four matrix elements are $3\times3$ matrices proportional to the identity.

In general, one can have an $S$-matrix that not only couples orbital angular momentum states but also flavor states and/or spin state. For example, consider a spin-singlet $\Lambda\Lambda$ system in a S-wave. The ground states of this channel, the H-dibaryon, has been observed to be a bound state for unphysical values of $m_\pi$~\cite{ Beane:2010hg, Beane:2011xf, Inoue:2011ai,Inoue:2010es}. In flavor space $\Lambda\Lambda$ also mixes with the $I=0$ $\Xi N$ and $\Sigma\Sigma$ channels. For low-energies, the positive parity $J$=0 $S$-matrix can be approximated as a $3\times 3$ matrix in flavor space. The $\Lambda\Lambda$ ground state must be an spin-singlet state due to the Pauli-exclusion principle, but the spin of the $\Xi N$ is not constrained by symmetry considerations. Therefore,  the ${^1}P_1$ and ${^3}P_1$ $\Xi N$ states mix and the corresponding $J=1$ $S$-matrix, which can be approximated to be a $6\times 6$ matrix, has nonzero elements coupling these two channels. For sufficiently high energies, the ${^3}P_1$-${^3}F_1$ mixing of the $\Xi N$ state may in general not be neglected. Furthermore, although the $S$-matrix does not couple ${^1}S_0$ and ${^1}P_1$ $\Xi N$ states, these may in general mix in a finite volume~\cite{Fu:2011xz}.   

Even though the scattering amplitude may in general not be diagonal in spin, spin is conserved in the infinite volume loops, $\mathcal{G}^{\infty}$. This is a consequence of the fact that the single particle propagators are diagonal in helicity. This also explains why spin is conserved in the finite volume loops, $\mathcal{G}^{V}$. The only difference between  $\mathcal{G}^{V}$ and  $\mathcal{G}^{\infty}$ is that the momenta of the intermediate particles is discretized for the former but continuous for the latter. This results in partial wave mixing in a finite volume. This is in agreement with what has previously been found for systems with nonzero spin~\cite{Bernard:2008ax, Gockeler:2012yj,Ishizuka:2009bx,  Briceno:2013lba} and will be reviewed in Sec.~\ref{sec:loop}.

The finite volume spectrum can be obtained from the poles of the sum of all amputated $\textbf{2}\rightarrow\textbf{2}$ finite volume diagrams, $\mathcal{M}^V,$\footnote{The poles of this object satisfy the same quantization condition as those of the finite volume correlation function~\cite{Kim:2005gf, Hansen:2012tf, Briceno:2012yi, Li:2012bi}. } which is represented in Fig.~\ref{fig:QC}. This is the analogous finite volume object to the infinite volume scattering amplitude, and it asymptotes to $\mathcal{M}$ as the volume is taken to infinity. This object satisfies the following matrix, summation equation
\begin{eqnarray}
\label{FV_prop}
i\mathcal{M}^V&=&-i{K}^*_{\text{rel}}
+i{K}^*_{\text{rel}}\mathcal{G}^{V}\mathcal{M}^V, 
\end{eqnarray}
where $\mathcal{G}^{V}$ is the finite volume s-channel loop, and in particular the matrix elements of the second term in the equation above can be written as
\begin{eqnarray}
i[{K}^*_{\text{rel}}\mathcal{G}^{V}\mathcal{M}]_{ab}
&\equiv&  \nn\\
&&\hspace{-2.5cm}
\frac{1}{L^3}\sum_{\bm{q}}\int\frac{dq^{0}}{2\pi}[{K}_{\text{rel}}({p}_f,q)]_{aj}\Delta_{j}(q)[\mathcal{M}(q,{p}_i)]_{jb}.
\label{FV_prop2}
\end{eqnarray}

As thoroughly discussed in Ref.~\cite{Kim:2005gf} for the spin-singlet, single channel scenario, the only power law finite volume corrections of $\mathcal{G}^{V}$ arise from the pole structure of the intermediate two-particle propagator. Therefore, the difference between this loop and the infinite volume counterpart, $\delta\mathcal{G}^V\equiv \mathcal{G}^V-\mathcal{G}^\infty$, depends on the on-shell momentum. The on-shell condition fixes the magnitude of the momentum running through the kernels but not its direction. Therefore it is convenient to decompose the product of the kernels and $\delta\mathcal{G}^V$ into spherical harmonics. These depend not only on the directionality of the intermediate momentum but also on those of the incoming and outgoing momenta. In Refs.~\cite{Hansen:2012tf, Briceno:2012yi, Li:2012bi} it was demonstrated that this persists to be true for coupled channel systems with $S=0$ or $S=1$. In fact, this observation is independent of the spin structure of the system of interest and the number of channels. For arbitrary number of channels one may simply upgrade $\delta\mathcal{G}^V$ to be not just a matrix in the spherical harmonic space but also a matrix in the open channels. If the system has non-zero spin, then it is convenient to represent the kernel and $\delta\mathcal{G}^V$ not as matrices in orbital angular momentum but rather total angular momentum. Just like the scattering amplitude, the kernel is diagonal in the total angular momentum. As a result, its matrix elements can be written in the same form as the scattering amplitude, Eq.~\ref{MJMlsbasis}. A derivation of $\delta\mathcal{G}^V$ is presented in Sec.~\ref{sec:loop}.

Having upgraded these objects to infinite dimensional matrices in $J$ and the space of open channels, it is easy to see that the poles of Eq.~\ref{FV_prop}  satisfy
\begin{eqnarray}
\det~[\mathcal{M}^{-1}+\delta \mathcal{G}^V]&=&\nn\\
&&\hspace{-1.5cm}{\det}_{\rm{oc}}\left[{\det}_{{lSJm_J}}~[\mathcal{M}^{-1}+\delta \mathcal{G}^V]\right]=0,
\label{eq:QC}
\end{eqnarray}
where the determinant ${\det}_{\rm{oc}}$ is over the N open channels and the determinant $\det_{{lSJm_J}}$ is over the $|lS,Jm_J\rangle$ basis, and both $\mathcal{M}$ and $\delta \mathcal{G}^V$ functions are evaluated on the on-shell value of the momenta, Eq.~\ref{momentumcc}. The matrix elements of $\delta\mathcal{G}^V$ for the $jth$ channel are defined as
\begin{widetext}
\begin{eqnarray}
&& \left[\delta\mathcal{G}^V_j\right]_{Jm_J,lS;J'm_{J'},l'S'}=\frac{ik^*_j\delta_{SS'}}{8\pi E^*}n_j\left[\delta_{JJ'}\delta_{m_Jm_{J'}}\delta_{ll'} +i\sum_{l'',m''}\frac{(4\pi)^{3/2}}{k^{*{l''}+1}_j}c_{l''m''}^{\mathbf{d}}(k^{*2}_j;{L}) \right.
\nn\\
&&\hspace{-1cm} \qquad \qquad \qquad \qquad ~ \left .  \times \sum_{m_l,m_{l'},m_{S}}\langle lS,Jm_J|lm_l,Sm_{S}\rangle \langle l'm_{l'},Sm_{S}|l'S,J'm_{J'}\rangle \int d\Omega~Y^*_{ l,m_l}Y^*_{l'',m''}Y_{l',m_{l'}}\right],
\label{eq:deltaGPBCs}
\end{eqnarray}
\end{widetext}
and the function $c^{\textbf{d}}_{lm}$ is defined as
\begin{eqnarray}
\hspace{1cm} 
c^\mathbf{d}_{lm}(k^{*2}_j; {L})
&=&\frac{\sqrt{4\pi}}{\gamma L^3}\left(\frac{2\pi}{L}\right)^{l-2}\mathcal{Z}^\mathbf{d}_{lm}[1;(k^*_j {L}/2\pi)^2],\nn\\
    \label{eq:clm}
\mathcal{Z}^\mathbf{d}_{lm}[s;x^2]
&=& \sum_{\mathbf r \in \mathcal{P}_{\mathbf{d}}}\frac{|\mathbf{r}|^lY_{l,m}(\mathbf{r})}{(r^2-x^2)^s},\label{eq:clm}
\end{eqnarray} 
where $\gamma=E/E^*$, the sum is performed over $\mathcal{P}_{\mathbf{d}}=\left\{\mathbf{r}\in \textbf{R}^3\hspace{.1cm} | \hspace{.1cm}\mathbf{r}={\hat{\gamma}}^{-1}(\mathbf m-\alpha_j \mathbf d) \right\}$, $\textbf{m}$ is a triplet integer, $\mathbf d$ is the normalized boost vector $\mathbf d=\mathbf{P}L/2\pi$, $\alpha_j=\frac{1}{2}\left[1+\frac{m_{j,1}^2-m_{j,2}^2}{E^{*2}}\right]$~\cite{Davoudi:2011md, Fu:2011xz, Leskovec:2012gb}, and $\hat{\gamma}^{-1}\textbf{x}\equiv{\gamma}^{-1}\textbf{x}_{||}+\textbf{x}_{\perp}$,  with $\textbf{x}_{||}(\textbf{x}_{\perp})$ denoting the $\textbf{x}$ component that is parallel(perpendicular) to the total momentum, $\textbf{P}$. Details regarding the representation of the s-channel loops as matrices in angular momentum are shown in Sec.~\ref{sec:loop}. 

In deriving the result, PBCs have been assumed on the spatial extents of the lattice. The boundary conditions of the system are encoded in the form of the $\mathcal{Z}$ functions. References~\cite{Agadjanov:2013kja, Briceno:2013hya} derived these for systems with nonzero momenta, arbitrary masses and twisted boundary conditions. For completeness, Sec.~\ref{sec:SpinFVTBCs} includes the result in the presence of arbitrary twist and asymmetry volumes. As discussed above, it is evident from Eq.~\ref{eq:deltaGPBCs} that $\delta \mathcal{G}^V$ is diagonal in spin, although the scattering amplitude may in general not be. Due to the reduction of rotational symmetry, $\delta \mathcal{G}^V$ mixes different orbital angular momentum states and consequently different $J$ states, as expected. For example, for systems with  $\textbf{d}=\{(0,0,0)$, $(0,0,n)$, $(n,n,0)$, $(n,n,n)$, $(n,m,0)$, $(n,n,m)$, $(n,m,p)\}$, or any cubic rotation of these, the symmetry point groups are the double cover of the octahedral ($\mathrm{O}^\mathrm{D}_h$)  and the dicyclic groups $\mathrm{Dic}_4$, $\mathrm{Dic}_2$, $\mathrm{Dic}_3$, $\mathrm{C}_4$, $\mathrm{C}_4$ \& $\mathrm{C}_2$, respectively. Table~\ref{table:irreps} lists the decomposition of the irreducible representations (irreps) of these three groups onto continuum states that have overlap with both half-integer and integer spin systems up to $J=4$~\cite{Luscher:1986pf, Luscher:1990ux, Feng:2004ua, Dresselhaus, Thomas:2011rh, Gockeler:2012yj, Moore:2005dw, Moore:2006ng, Dudek:2010wm, Dudek:2012gj}.

 As was mentioned in Sec.~\ref{sec:Intro}, the master equation presented here, Eq.~\ref{eq:QC}, is consistent with all previous results. The most general multichannel result for scalars was presented in Refs.~\cite{Hansen:2012tf, Briceno:2012yi}. If one restricts the total spin of all of the available channels to be exactly zero ($S=0$ in Eq.~\ref{eq:deltaGPBCs}), then Clebsch-Gordan coefficients are all replaced with Kronecker delta functions setting orbital and total angular momenta equal to each one and one recovers the result of these references. If $S=1/2$ in Eq.~\ref{eq:deltaGPBCs} and furthermore restricts there to be only a single channel, one recovers the result of Ref.~\cite{Bernard:2008ax, Gockeler:2012yj}. If one allows for arbitrary numbers of channels with $S=1/2$ then one arrives at the result of Ref.~\cite{Li:2012bi}. Allowing for a single channel with $S=1$ and restricting the energies to be relativistic, i.e., $\gamma\approx1$, one arrives at the two-nucleon result shown in Refs.~\cite{Ishizuka:2009bx, Briceno:2013lba, Briceno:2013bda, Briceno:2013pda}.

Although what is presented here is the master equation describing the full finite volume spectrum for arbitrary two-body systems, in practice one needs to reduce the master equation onto the quantization condition of the irreps of the system of interest. For systems with PBCs, there has been a great deal of effort in reduction of these master equation for a wide variety of scenarios ~\cite{Luscher:1986pf, Luscher:1990ux, Luu:2011ep, Dudek:2010wm, Rummukainen:1995vs, Feng:2004ua, Bernard:2008ax, Leskovec:2012gb, Thomas:2011rh, Dudek:2012gj, Gockeler:2012yj}. References~\cite{Dudek:2010wm, Thomas:2011rh, Dudek:2012gj} demonstrate how to decompose the master equation for integer-spin systems for the irreps of the symmetry point groups corresponding to ${\textbf{d}}=\{(0,0,0)$, $(0,0,n)$, $(0,n,n)$, $(n,n,n)$, $(n,m,0)$, $(n,n,m)\}$. References \cite{Bernard:2008ax, Gockeler:2012yj} contain the relations of the non-vanishing $c_{lm}$ functions as well as the basis vectors for $S=1/2$ systems with ${\textbf{d}}=\{(0,0,0)$, $(0,0,n)$, $(0,n,n)$, $(n,n,n)\}$. 
%
 
\begin{table}
\begin{center}

\subtable[]{ \label{table:irrepsa}
\begin{tabular}{lc}
\hline\hline
$J^P$&$\mathrm{O^D_h}$ \\\hline
$0^\pm$&${A}_1^\pm$\\
 $\frac{1}{2}^\pm$&$G_1^\pm$\\
 $1^\pm$&${T}_1^\pm$\\
$\frac{3}{2}^\pm$&$H^\pm$\\
$2^\pm$&$E^\pm\oplus T_2^\pm$ \\
$\frac{5}{2}^\pm$&$G_2^\pm \oplus H^\pm  $\\
$3^\pm$&$A_2^\pm\oplus T_1^\pm \oplus T_2^\pm$\\
$\frac{7}{2}^\pm$&$ G_1^\pm \oplus G_2^\pm \oplus H^\pm $\\
$4^\pm$&$A_1^\pm\oplus E ^\pm\oplus T_1^\pm \oplus T_2^\pm$\\
\hline\hline
\end{tabular}

} \subtable[]{ \label{table:irrepsb}
\begin{tabular}{lllllllllll}
\hline\hline
$|\lambda|^{\tilde{\eta}}$ & $\mathrm{Dic}_4$ & $\mathrm{Dic}_2$& $\mathrm{Dic}_3$  & $\mathrm{C_{4}}$ & $\mathrm{C_{2}}$\\
  \hline
  $0^+$         & $A_1$            & $A_1$            & $A_1$            & $A$ & $A$ \\
  $0^-$         & $A_2$            & $A_2$            & $A_2$            & $B$ & $A$ \\
  $\frac{1}{2}$ & $E_1$            & $E$            & $E_1$              & $E$ & $2B$ \\
  $1$           & $E_2$            & $B_1	\oplus B_2$			& $E_2$              & $A \oplus B$ & $2A$  \\
  $\frac{3}{2}$ & $E_3$            & $E$             & $B_1 \oplus B_2$  & $E$ & $2B$ \\
  $2$           & $B_1 \oplus B_2$ & $A_1 \oplus A_2$& $E_2$             & $A \oplus B$ & $2A$ \\
  $\frac{5}{2}$ & $E_3$            & $E$            & $E_1$              & $E$ & $2B$ \\
  $3$           & $E_2$            & $B_1 \oplus B_2$ & $A_1 \oplus A_2$ & $A \oplus B$& $2A$  \\
  $\frac{7}{2}$ & $E_1$            & $E$            & $E_1$              & $E$ & $2B$ \\
  $4$           & $A_1 \oplus A_2$ & $A_1 \oplus A_2$  & $E_2$            & $A \oplus B$& $2A$\\
\hline\hline
\end{tabular}
}
\caption{ \label{table:irreps} (a)~The decomposition of the irreps of the SO(3) group up to $J=4$ in terms of the irreps of the $\mathrm{O^D_h}$~\cite{Luscher:1986pf, Luscher:1990ux, Mandula:1983wb, Johnson:1982yq, Basak:2005aq}.
(b) The decomposition of the helicity states to the irreps of five of the little groups of $\mathrm{O^D_h}$: $\mathrm{Dic}_4$, $\mathrm{Dic}_2$, $\mathrm{Dic}_3$, $\mathrm{C}_4$ \& $\mathrm{C}_2$~\cite{Moore:2005dw, Moore:2006ng, Dudek:2010wm, Thomas:2011rh, Dudek:2012gj}. $\lambda$ labels the helicity of the state and $\tilde{\eta}=P(-1)^J$, where $P$ is the parity of the state. 
}
\end{center}
\end{table}

\subsection{Relativistic finite volume loop with spin \label{sec:loop}}

In the absence of weak interactions, the free two-particle propagators are diagonal in the open channels. That is to say, in the absence of two-body interactions the different channels would not mix. This is depicted in Fig.~\ref{fig:FVprop}. As a consequence, it is only necessary to investigate the structure of the s-channel loop appearing in one of the open channels. Therefore, to alleviate some of the strenuous notation that is necessary when discussing coupled channel systems, we will momentarily drop the $j$ subscript that explicitly reminds the reader that the $jth$ channel of potentially infinitely many open channels is being discussed.

\begin{figure*}[t]
\begin{center}
\includegraphics[scale=0.4]{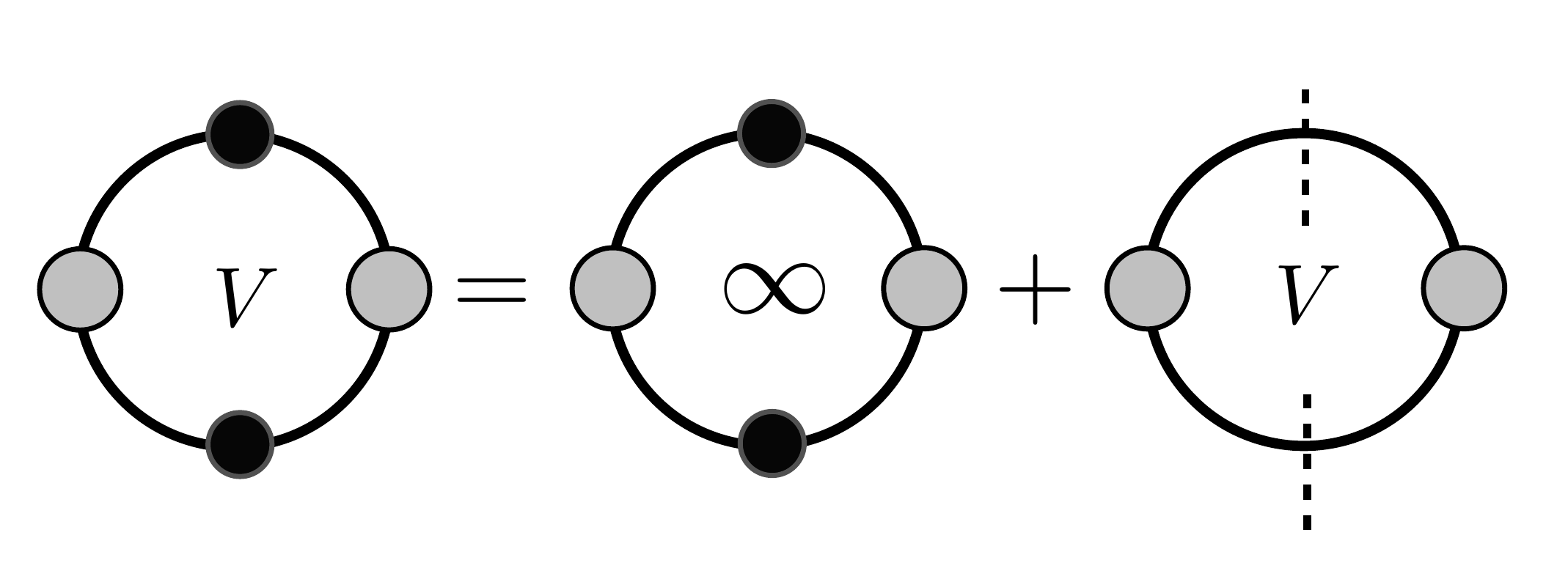}
\caption{ Shown is the close-up of a generic finite volume loop appearing in the determination of the quantization condition, Eq.~\ref{eq:QC}, which is determined from the poles of the $\mathcal{M}^V$, defined in Eq.~\ref{FV_prop} and pictorially depicted in Fig.~\ref{fig:QC}.  The finite volume loop can always be set equal to its infinite volume counterpart up to finite volume correction. In Sec.~\ref{sec:loop}, it is shown that this correction can be written as a product of infinite-dimensional matrices that solely depend on the on-shell momenta of the intermediate particles in the loop.}\label{fig:dGV}
\end{center}
\end{figure*}

In order to evaluate the sum depicted in Fig.~\ref{fig:dGV}, it is convenient to upgrade the kernel onto a matrix in spin. Depending on the nature of the the particles of interest, bosonic vs. fermionic, the dimensionality of the single particle propagator will differ. Nevertheless, the single particle poles will satisfy the relativistic dispersion relation for all particles. Alternatively, one may always perform a field redefinition to assure the propagator of bosonic and fermionic field have the same dimensions, and in doing so one can define the residues appearing in Eq.~\ref{bareprop} to be equal to one when the particles go on-shell. This allows one to write the difference between the first term of the finite volume loop depicted in Fig.~\ref{fig:dGV} and its infinite volume counterpart in the following form
\begin{widetext}
\begin{eqnarray}
\label{loop1}
i\delta G^{V}&\equiv&n\left[~\frac{1}{L^3}\sum_{\mathbf{q}}\hspace{-.4cm}\int~\right]\int\frac{dq^0}{2\pi} \frac{K_{\text{rel}}({p}_f,{q})~K_{\text{rel}}({q},{p}_i)~z_{1}(P-q)~z_{2}(q)}{[(q-P)^2-m_{1}^2+i\epsilon][q^2-m_{2}^2+i\epsilon]},
\end{eqnarray}
\end{widetext}
where the kernels are being represented as matrices in helicity and the dependence on the total four-momentum $P$ is being suppressed, and the following notation has been introduced
\begin{eqnarray}
\left[~\frac{1}{L^3}\sum_{\mathbf{q}}\hspace{-.4cm}\int~\right]\equiv\left(\frac{1}{L^3}\sum_{\mathbf{q}}-\int\frac{d\textbf{q}}{(2\pi)^3}\right).
\end{eqnarray}
More explicitly, the product of the two kernels in Eq.~\ref{loop1} should be interpreted as
\begin{eqnarray}
K_{\text{rel}}({p}_f,{q})~K_{\text{rel}}({q},{p}_i)
&=&\nn\\
&&\hspace{-4cm}\sum_{\lambda_1,\lambda_2}
\hat K_{\text{rel}}({p}_f,{q})
|\mathbf{q-P},s_1\lambda_1\rangle
\otimes|-\textbf{q},s_2 \lambda_2\rangle\nn\\
&&\hspace{-3cm}
\langle-\textbf{q},s_2 \lambda_2|\otimes
\langle\mathbf{q-P},s_1\lambda_1|
\hat K_{\text{rel}}({q},{p}_i).
\label{prodKhelicity}
\end{eqnarray}
This is to emphasize that the single particle propagators are diagonal in helicity. Because there is a complete set of states between the two kernels, one can always perform a unitary transformation to represent this product in an alternative basis.

In general, the kernel is a function of volume, but since the c.m. energy is restricted to satisfy $m_{j,1}+m_{j,2}~\lsim E^*\ll  E^*_{th}$ the intermediate particles appearing in the kernel, Fig.~\ref{fig:kernel}, cannot all simultaneously go on-shell. Therefore, one can show using Poisson's resummation formula,  
\begin{eqnarray}
\label{poisson}
\left[~\frac{1}{L^3}\sum_{\mathbf{q}}\hspace{-.4cm}\int~\right] f(\textbf{q})=  
\sum_{\textbf{n}\neq 0}    \int \frac{d\mathbf{q}}{(2\pi)^3} f(\textbf{q})~{e^{iL \mathbf{n}\cdot \mathbf{q}}}. \nn
\end{eqnarray}
 that this leads to exponentially small deviations from the infinite volume kernel. By neglecting these corrections, the result discussed here holds for volumes satisfying $m_\pi L\gg 1$. We will also neglect terms in $\delta G^V$ that are exponentially suppressed with the mass of any of the two-particles in the given channel since $\mathcal{O}(e^{-m_{i} L})\leq \mathcal{O}(e^{-m_\pi L})$. These corrections have been previously determine for $\pi\pi$~\cite{Bedaque:2006yi} and $NN$ systems~\cite{Sato:2007ms} in an S-wave, as well as the $\pi\pi$ system in a P-wave in Ref.~\cite{Chen:2012rp, Albaladejo:2013bra}. 
 
The identification of the power law volume dependence of this function is most readily done by rewriting the summand in terms of the c.m. of coordinates. To do this the notation used in Ref.~\cite{Kim:2005gf} will be used
\begin{eqnarray}
\omega_{q,i}=\sqrt{|\textbf{q}|^2+m_i^2}.
\end{eqnarray}
The Lab frame coordinates $\textbf{q}=({q}_{||},{q}_\perp)$ and $\omega_{q,i}$ appearing in the summand above can be transformed to c.m. coordinates $\textbf{q}^*=({q}^*_{||},{q}^*_\perp)$ and $\omega_{q,i}^*=\sqrt{{q}^{*2}+m_i^2}$ using the standard Lorentz transformations 
\begin{eqnarray}
\omega_{q,i}^*&=&\gamma(\omega_{q,i}-\beta {q}_{||}), \hspace{.2cm}\nn\\
{q}^*_{||}&=&\gamma({q}_{||}-\beta \omega_{q,i}),\hspace{.8cm} 
{q}^*_{\perp}={q}_{\perp},
\end{eqnarray}
where $\gamma = \frac{E^*}{E},~\beta=\frac{P}{E}$. Using these relations, writing the functional form of the kernels in the c.m. frame as $K_{\text{rel}}^{*}$, and neglecting exponentially suppressed corrections, Eq.~\ref{loop1} can be rewritten as~\cite{Kim:2005gf,Briceno:2013rwa}

\begin{widetext}
\begin{eqnarray} 
i\delta G^{V}
&=&-in\left[~\frac{1}{L^3}\sum_{\mathbf{q}}\hspace{-.4cm}\int~\right]\frac{1}{2E^*}\frac{\omega_{q,2}^*}{\omega_{q,2}}
\frac{{K}^*_{\text{rel}}(\mathbf{p}_f^*,\mathbf{q}^*){K}^*_{\text{rel}}(\mathbf{q}^*,\mathbf{p}_i^*)~z_{1}^*(q^*)~z_{2}^*(q^*)}{k^{*2}-q^{*2}+i\epsilon}\left(\frac{E^*+\frac{m_1^2-m_2^2}{E^*}+2\omega^*_{q,2}}{4\omega^*_{q,2}}\right).
\end{eqnarray}
\end{widetext}
where $k^*$ is the on-shell c.m. momentum and satisfies Eq.~\ref{momentumcc}. In general the kernel will also depend on $\omega_{q,i}^*$, but since this is itself a function of $\textbf{q}^*$ the explicit dependence on $\omega_{q,i}^*$ has been suppressed.

By restricting themselves to the scalar sector, Kim, Sachrajda, and Sharpe showed that this summation can be represented as a product of infinite-dimensional matrices in orbital angular momentum~\cite{Kim:2005gf}. This result can be recovered by decomposing the product of the two kernels into spherical components,
\begin{eqnarray} K_{\text{rel}}^*(\mathbf{p}_f^*,\mathbf{q}^*)K_{\text{rel}}^*(\mathbf{q}^*,\mathbf{p}_i^*)&=&\nn\\
&&\hspace{-1cm}\sum_{l,m}
f_{lm}(q^*){\sqrt{4\pi}}{q^{*l}}Y_{lm}(\hat{q}^*).
\label{eq:flm}
\end{eqnarray}
The function $f_{lm}$ is defined as to satisfy this equation and its definition in terms of the spherical decomposition of the kernels is easy to write down. Using this function one finds that $\delta G^V$ can be written as~\cite{Kim:2005gf}
\begin{eqnarray} 
i\delta G^{V} 
&=&n\left(\frac{k^*f_{00}(k^*)}{8\pi E^*}+\frac{i}{2 E^*}\sum_{l,m}f^*_{lm}(k^*)c^{\textbf{d}}_{lm}(k^{*2})\right),\nn
\end{eqnarray}
where $c^{\textbf{d}}_{lm}$ has been defined in Eq.~\ref{eq:clm}. This expression holds for arbitrary spin systems. Section~\ref{sec:helicityLS} showed that one can decompose any object that is diagonal in angular momenta, such as the scattering amplitude and the kernel, in the $lS$ basis with matrix elements shown in Eq.~\ref{MJMlsbasis}. Using these expressions along with Eqs.~\ref{prodKhelicity} \& \ref{eq:flm} one finds\footnote{For further details see Ref.~\cite{Briceno:2013rwa}.}
\begin{eqnarray} 
i\delta G^{V}&=&
-i[K_{\text{rel}}^{*}]^{Jm_J}_{lS}(\delta \mathcal{G}^{V})_{Jm_J,lS;J'm_{J'},l'S'}[K_{\text{rel}}^{*}]^{J'm_J'}_{l'S'}\nn\\
&=&
[-iK_{\text{rel}}^{*}](i\delta \mathcal{G}^{V})[-iK_{\text{rel}}^{*}],
\label{eq:dGV}
\end{eqnarray}
where we have suppressed the indices of the incoming and outgoing state in the loop, $[K_{\text{rel}}^{*}]^{Jm_J}_{lS}$ denote the on-shell kernels, and the matrix elements $(\delta \mathcal{G}^{V})_{Jm_J,lS;J'm_{J'},l'S'}$ are defined in Eq.~\ref{eq:deltaGPBCs}. Equation~\ref{eq:dGV} shows that the difference between the finite volume and infinite volume loops can be represented in a matrix representation of functions that only depend on the on-shell momenta. Having shown this for a single channel allows one to quickly derive the relation for an arbitrary numbers of channels. In general one could have one species, $``a"$, going into the loop and another one, $``b"$, outgoing. By upgrading all the objects appearing in Eq.~\ref{eq:dGV} in the space of channels, one finds, 
\begin{eqnarray} 
i\delta G^{V}_{ba}&=&[-iK_{\text{rel}}^{*}]_{bj}(i\delta \mathcal{G}^{V}_j)[-iK_{\text{rel}}^{*}]_{ja}
\label{eq:dGVj},
\end{eqnarray}
where the intermediate $j$-index is summed over all open channels. By utilizing this relation along with the definition of the infinite volume scattering amplitude in terms of the kernel, Eq.~\ref{IV_prop}, and the definition of $\mathcal{M}^V$, Eq.~\ref{FV_prop}, one arrives at the quantization condition, Eq.~\ref{eq:QC}.


\section{Two-particle multichannel systems with spin with TBCs in asymmetric volumes\label{sec:SpinFVTBCs}}
\subsection{Cubic volumes  \label{sec:SpinFVTBCssym}}
In the derivation of the master equation shown in Eq.~\ref{eq:QC}, periodic boundary condition on the spatial extent of the cubic volume have been assumed. The periodicity constraint is encoded in the expression for the $\mathcal{Z}$ functions shown in Eq.~\ref{eq:clm}, and this is generally true for arbitrary boundary conditions. As discussed in Sec.~\ref{sec:Intro}, TBCs require that fields are proportional to their images up to an overall phase. Therefore, particle $``1"$ in the $jth$ channel will have a free discretized momenta satisfying $\textbf{p}_{j,1}=\frac{2\pi\textbf{n}}{L}+\frac{\bm{\phi}_{j,1}}{L}$, where $\bm{\phi}_{j,1}$ is the three-dimensional phase for that particle. Each particle in each channel could have an overall different phase which, when thinking of LQCD calculations, is dictated by the quark content of the hadron in mind. As a consequence, the total momentum of the systems will be shifted to $\textbf{P}=\frac{2\pi\textbf{d}}{L}+\frac{\bm{\phi}_{j,1}+\bm{\phi}_{j,2}}{L}$. Although, it may naively seem that the total momentum would depend on the channel considered, it is easy to convince oneself that for coupled channel systems ${\phi}_{j,1}+\bm{\phi}_{j,2}$ is a conserved quantity, since antiquark fields satisfy 
\begin{eqnarray}
\bar{\psi}(\mathbf{x}+\mathbf{n}{{L}})=e^{-i{\bm{\theta}} \cdot \mathbf{n}}\bar{\psi}(\mathbf{x}).
\end{eqnarray}

Having defined the total momentum, the relationship between the total energy and the c.m. energy remain unchanged, and the c.m. on-shell momenta for the $jth$ channel still satisfies Eq.~\ref{momentumcc}. The only part of the master equation that is modified is the finite volume function, Eq.~\ref{eq:deltaGPBCs}. One finds that the $c_{lm}$ and $\mathcal{Z}$ functions with arbitrary twist for a cubic volume is~\cite{Bernard:2010fp, Doring:2011vk, Ozaki:2012ce, Agadjanov:2013wqa, Briceno:2013hya}, 
\begin{eqnarray}
c^{\textbf{d},\bm{\phi}_{j,1},\bm{\phi}_{j,2}}_{lm}(k^{*2};{L})
\ &=&\ \frac{\sqrt{4\pi}}{\gamma {L}^3}\left(\frac{2\pi}{{L}}\right)^{l-2}\nn\\
&&\hspace{-.1cm}\times\mathcal{Z}^{\mathbf{d},\bm{\phi}_{j,1},\bm{\phi}_{j,2}}_{lm}[1;(k^*{L}/2\pi)^2],
\label{clmTBCs}\\
\mathcal{Z}^{\mathbf{d},\bm{\phi}_{j,1},\bm{\phi}_{j,2}}_{lm}[s;x^2]
\ &=&\ \sum_{\mathbf r \in \mathcal{P}_{\mathbf{d}}^{\bm{\phi}_1,\bm{\phi}_2}}
\frac{ |{\bf r}|^l \ Y_{l,m}(\mathbf{r})}{(\mathbf{r}^2-x^2)^s},
\label{ZlmTBCs}
\end{eqnarray}
where $ \mathcal{P}_{\mathbf{d}}^{\bm{\phi}_1,\bm{\phi}_2}=\left\{\mathbf{r}\in \textbf{R}^3\hspace{.1cm} | \hspace{.1cm}\mathbf{r}={\hat{\gamma}}^{-1}(\mathbf m-\alpha_j \mathbf d +\frac{\bm{\Delta}^{(j)}}{2\pi})\right\}$, where $\textbf{m}$ is a triplet integer, $\bm{\Delta}^{(j)}=-(\alpha_{j}-\frac{1}{2})(\bm{\phi}_{j,1}+\bm{\phi}_{j,2})+\frac{1}{2}(\bm{\phi}_{j,1}-\bm{\phi}_{j,2})$. Just as before, $\hat{\gamma}^{-1}\textbf{x}\equiv{\gamma}^{-1}\textbf{x}_{||}+\textbf{x}_{\perp}$,  with $\textbf{x}_{||}(\textbf{x}_{\perp})$ denoting the $\textbf{x}$ component that is parallel(perpendicular) to the total momentum, $\textbf{P}$.

With this, one arrives at the conclusion that the quantization condition for the spectrum of a two-particle multi-channel system with TBCs can still be written as Eq.~\ref{eq:QC}, where the matrix elements of $\delta\mathcal{G}^V$ for the $jth$ channel can be obtained by replacing $c_{l''m''}^{\mathbf{d}}(k^{*2}_j;{L})$ with $c^{\textbf{d},\bm{\phi}_{j,1},\bm{\phi}_{j,2}}_{l''m''}(k^{*2}_j;{L})$ in Eq.~\ref{eq:deltaGPBCs}. One important observation is that if the two-particles are degenerate and they have the same twist then twisting will have no overall impact in the c.m. spectrum. Therefore, one may not gain any additional information for systems like $\pi^+\pi^+$ or $pp$ using TBCs. Furthermore, if isospin is exact and the twist on the up and down quarks is the same, this will give the same $pn$ c.m. spectrum  as if it was at rest and untwisted. Reference~\cite{Briceno:2013hya} investigated the implication of the determination of the deuteron binding energy when using asymmetric twists on the up and down quarks, and found that by introducing an overall twist $\bm{\phi}_{p}=-\bm{\phi}_{n}=(\pi/2,\pi/2,\pi/2)$ finite volume artifacts of the deuteron binding energies can be reduced from $\sim e^{-\kappa L}/L$ to $\sim e^{-2\kappa L}/L$, where $\kappa$ is infinite volume binding momentum of the deuteron. 

Another important remark is that when introducing an arbitrary twist, partial wave mixing can be a subtle matter. This is due to the rich structure of the $c_{lm}$'s in Eq.~\ref{clmTBCs}. For instance, for the scenario discussed in Ref.~\cite{Briceno:2013hya} the S-wave deuteron channel not only has physical mixing with the $^3D_1$, but in general will have finite volume mixing with the $^3P_0$, $^3P_1$, $^3P_2$, $^3D_2$ and $^3D_3$ channels, as well as higher partial waves, even when the up and down quark masses are exactly degenerate. 
\subsection{Asymmetric volumes \label{sec:SpinFVTBCsasym}}
References~\cite{Li:2003jn, Detmold:2004qn, Feng:2004ua} demonstrated how the \textit{L\"uscher} method can be generalized for asymmetric volumes. Adopting the notation introduced in these references, let $L$ be the spatial extent of the z-axis and $\eta_i$ be the asymmetric factor of the $ith$ axis, i.e., $L_x=\eta_xL$ and  $L_y=\eta_yL$. In evaluating the finite volume loop in the previous section, Eq.~\ref{loop1}, one must make the following replacement, 
\begin{eqnarray}
\frac{1}{L^3}\sum_{\mathbf{q}}
\rightarrow\frac{1}{\eta_x\eta_y L^3}\sum_{\mathbf{q}}~,
\end{eqnarray}
which leads to an overall factor of $({\eta_x\eta_y})^{-1}$ in the $c_{lm}$ functions, Eq.~\ref{clmTBCs}. Furthermore, the free particle momenta are altered. Let $\bm{\chi}$ be an arbitrary three-dimensional vector. By introducing the notation $\tilde{\bm{\chi}}=(\chi_x/\eta_x,\chi_y/\eta_y,\chi_z)$, one can readily find that the $``1"$ in the $jth$ channel has a free momentum of $\textbf{p}_{j,1}=\frac{2\pi\tilde{\textbf{n}}}{L}+\frac{\tilde{\bm{\phi}}_{j,1}}{L}$, where $\textbf{n}$ is a integer triplet and ${\bm{\phi}}_{j,1}$ is the particle's twist. With these pieces one may arrive at the most general form of the $c_{lm}$ and $\mathcal{Z}$ functions with arbitrary twist for an asymmetric volume
\begin{eqnarray}
c^{\textbf{d},\bm{\phi}_{j,1},\bm{\phi}_{j,2}}_{lm}(k^{*2};{L};\eta_x,\eta_y)
\ &=&\ \frac{\sqrt{4\pi}}{\eta_x\eta_y\gamma {L}^3}\left(\frac{2\pi}{{L}}\right)^{l-2}\nn\\
&&\hspace{-2cm}\times\mathcal{Z}^{\mathbf{d},\bm{\phi}_{j,1},\bm{\phi}_{j,2}}_{lm}[1;(k^*{L}/2\pi)^2;\eta_x,\eta_y],~~
\label{clmasymTBCs}\\
\mathcal{Z}^{\mathbf{d},\bm{\phi}_{j,1},\bm{\phi}_{j,2}}_{lm}[s;x^2;\eta_x,\eta_y]
\ &=&\ \sum_{\mathbf r \in \mathcal{P}_{\mathbf{d};\eta_x,\eta_y}^{\bm{\phi}_1,\bm{\phi}_2;}}
\frac{ |{\bf r}|^l \ Y_{l,m}(\mathbf{r})}{(\mathbf{r}^2-x^2)^s},~~~
\label{ZlmasymTBCs}
\end{eqnarray}
where $ \mathcal{P}_{\mathbf{d};\eta_x,\eta_y}^{\bm{\phi}_1,\bm{\phi}_2}=\left\{\mathbf{r}\in \textbf{R}^3\hspace{.1cm} | \hspace{.1cm}\mathbf{r}={\hat{\gamma}}^{-1}(\tilde{\mathbf m}-\alpha_j \tilde{\mathbf d} +\frac{\tilde{\bm{\Delta}}^{(j)}}{2\pi})\right\}$, where $\textbf{m}$ is a triplet integer, $\tilde{\bm{\Delta}}^{(j)}=-({\alpha}_{j}-\frac{1}{2})(\tilde{\bm{\phi}}_{j,1}+\tilde{\bm{\phi}}_{j,2})+\frac{1}{2}(\tilde{\bm{\phi}}_{j,1}-\tilde{\bm{\phi}}_{j,2})$. In the limit that the total momentum and twist of the system vanishes, this result agrees with Refs.~\cite{Li:2003jn, Detmold:2004qn, Feng:2004ua}. The boost vector, $\tilde{\mathbf d}$, is defined to be equal to ${\mathbf P}L/2\pi$. It is important to note that in the limit where the twist angles of the two particles vanish, the boost vector for asymmetric volumes in general is not an integer triplet.

\section{Implication for baryon-baryon systems\label{sec:baryon2}}

As was discussed in the previous sections, the formalism presented here is universal and gives a mapping between the finite volume spectrum and the infinite volume scattering amplitude for arbitrary two-particle systems. A sector of physics where this formalism will have a clear and immediate impact is on the study of light nuclei and hyper nuclei from LQCD. This is a field that has received a great deal of excitement in recent years~\cite{Beane:2010hg, Inoue:2010es, Inoue:2011ai,Beane:2011xf, Beane:2012vq, Beane:2012ey,Yamazaki:2012hi,Beane:2011iw, Beane:2013br, Buchoff:2012ja, Francis:2013lva}. As was alluded to in the previous sections, these are systems with rather rich structure and with potential partial wave mixing in the infinite volume and/or several inelastic thresholds. For instance, the determination of hyperon-nucleon scattering phase shifts studied in Ref.~\cite{Beane:2012ey} was limited by the fact that this formalism was not known. Having the formalism in place, future calculations of these systems will no longer be restricted to the study of the ground state, where presumably only $S$-wave phase shifts are prevalent, but also scattering parameters will be able to be determined from excited states. The study in Ref.~\cite{Beane:2012ey} explicitly avoided coupled channels systems, e.g., $I=1/2$ $N\Sigma$-$N\Lambda$. Although this remains to be a computationally challenging problem, there is no formal restriction for determining not only scattering phase shifts but also mixing angles and thereby unfolding the rich structure of these systems. 
 
The need for performing calculations with multiple total momenta has been extensively advocated in the literature~\cite{Rummukainen:1995vs, Christ:2005gi, Kim:2005gf, Fu:2011xz, Bernard:2008ax, Gockeler:2012yj, Briceno:2013bda}. When boosting a given system its $c.m.$ energy in a finite volume is altered. This is evident from the quantization condition shown in Eq.~\ref{eq:QC}. The scattering amplitude only depends on c.m. coordinates, while the $c_{lm}$ functions, Eq.~\ref{eq:clm}, and consequently $\delta G^V$, Eq.~\ref{eq:deltaGPBCs}, depend on both the c.m. coordinate and total momenta of the system. Therefore the c.m. energies where Eq.~\ref{eq:QC} vanishes will in general differ for different boosts. This is extremely advantageous when trying to constrain the scattering amplitude from the finite volume spectrum, since it is at these energies where the scattering amplitude is determined. For coupled channel systems, boosting is a necessity~\cite{Hansen:2012tf, Briceno:2012yi}. For example, the $J=1$ $S=1$ matrix shown in Eq.~\ref{J1Smatrix} depends on three functions of energy, the S-wave and D-wave phase shifts and the mixing angle that couples these two channels. Therefore, from a single energy one can only constrain a linear combinations of these functions. Performing calculations with multiple boosts aids in disentangling these functions from the spectrum. In Ref.~\cite{Briceno:2013bda} it was shown just how to do this for the ${^3}S_1$-${^3}D_1$ two nucleon channel. An alternative tool for coupled channel systems is to perform calculations with twisted boundary conditions~\cite{Bernard:2010fp, Doring:2011vk, Ozaki:2012ce, Agadjanov:2013wqa, Briceno:2013hya} or asymmetric volumes~\cite{Li:2003jn, Detmold:2004qn, Feng:2004ua}.  

Furthermore, Refs.~\cite{Briceno:2013lba,Briceno:2013bda} went into great detail in demonstrating that the presence of partial wave mixing in the infinite volume could lead to an unexpectedly large effect in the boosted c.m. finite volume spectrum. For the deuteron channel, it was demonstrated that at the physical point these effects can lead to a $\sim50\%$ correction to the binding energies for moderate volumes of $L\sim9~\text{fm}$. This observation is expected to also hold for $S=1$ hyperon-nucleon/hyperon-hyperon systems.

In Ref.~\cite{Briceno:2013lba} it was assumed that isospin is exact, which in the infinite volume, where parity and total angular momentum are good quantum numbers, leads to spin conservation. For instance, this would suggest that ${^1}P_1$ and ${^3}P_1$ NN channels could not mix. In nature, up and down quark masses are not degenerate, and searches for experimental (e.g., see Refs.~\cite{Abegg:1988kx, Vigdor:1992pb, Abegg:1995qi,  Abegg:1998sg}) and theoretical consequences (see Ref.~\cite{Miller:2006tv} for a review on the topic) in the two-body sector of this reduction of symmetry are challenging. By performing calculations with non-degenerate up and down quark masses, future LQCD calculations will be able to further constrain the mixing between different spin channels.

 Although the discussion is focused on the baryon-baryon sector, this formalism will also be necessary for future studies of meson-meson or meson-baryon processes where one or both particles have spin. An example of such systems is the $J/\Psi$-$\phi$ scattering channel, which was recently studied in Ref.~\cite{Ozaki:2012ce} using TBCs.  This benchmark calculation determined the ${^1}S_0$ and ${^1}P_1$ $J/\Psi$-$\phi$ phase shifts using configurations with a lightest pion mass of $m_\pi=156$~MeV in hopes of finding evidence for the $Y (4140)$ resonance~\cite{Aaltonen:2009tz, Aaltonen:2011at}. In obtaining their result the authors have made two  two reasonable approximations. The first approximation refers to the fact that although the authors of Ref.~\cite{Ozaki:2012ce} accounted for the finite volume partial wave mixing of the ${^1}S_0$ and ${^1}P_1$ waves, they did not include effects due to physical mixing between the ${^1}P_1$, ${^3}P_1$ and ${^5}P_1$ waves in their analysis. This is expected to be a small contribution for non-relativistic systems, but in general the quantization condition presented in this work can be used to include such effects. The second approximation refers to the unstable nature of the $J/\Psi(1S)$ and/or $\phi(1020)$. Although Ref.~\cite{Roca:2012rx} quantitatively demonstrated that for a resonances such as the $\rho$, with a decay width of 147.8(9)~MeV~\cite{Beringer:1900zz}, one may not use two-body formalism presented here and used in Ref.~\cite{Ozaki:2012ce}, this formalism is expected to accurately describe the spectrum of a system including the $J/\Psi(1S)$ and/or the $\phi(1020)$. This is because their respective decay widths are 92.9(2.8)~keV and 4.26(4)~MeV~\cite{Beringer:1900zz} and their hadronic decays are in general suppressed by the OZI (Okubo-Zweig-Iizuka) rule~\cite{Okubo:1963fa,Zweig64, doi:10.1146/annurev.ns.20.120170.001445, Iizuka:1966fk}.

\section{Conclusion}
This paper presents the most general two-body finite volume formalism that gives the relationship between the finite volume spectrum and the infinite volume $\textbf{2}\rightarrow \textbf{2}$ scattering amplitude. The result holds for an arbitrary number of open two-body channels, with arbitrary masses, spin and momenta. The only restrictions is that the c.m. energy lies below the three-body inelastic threshold and that the spatial extent to the volume is significantly larger than the range of the interactions. It is evident from the result, that it is consistent with all previous two-body finite volume results~\cite{Luscher:1986pf, Luscher:1990ux, Rummukainen:1995vs, Li:2003jn, Detmold:2004qn, Feng:2004ua, He:2005ey, Liu:2005kr, Christ:2005gi, Kim:2005gf, Bernard:2008ax, Bour:2011ef, Davoudi:2011md, Leskovec:2012gb,  Gockeler:2012yj, Hansen:2012tf, Briceno:2012yi,  Li:2012bi, Guo:2012hv, Bernard:2010fp, Ishizuka:2009bx, Briceno:2013lba, Bernard:2010fp, Doring:2011vk, Ozaki:2012ce, Briceno:2013hya, Agadjanov:2013wqa}.

Section~\ref{sec:helicityLS} reviewed the basics of the construction of helicity states and their relation with the $lS$ basis~\cite{Jacob:1959at, :/content/aip/journal/jmp/4/4/10.1063/1.1703981, McKerrell64,  Chung:2007nn}.
Section~\ref{sec:SpinFVPBCs} presented a derivation of the quantization condition for multichannel systems with arbitrary spin, Eq.~\ref{eq:QC}, using generic aspects of relativistic quantum field theory. Sections~\ref{sec:SpinFVTBCssym} \& \ref{sec:SpinFVTBCsasym} presented the generalization of this result for systems with arbitrary TBCs in a cubic and asymmetric volume, respectively. Although the result is generic and independent of the nature of the particles of interest, Section~\ref{sec:baryon2} discussed the implication of this formalism for two-baryon systems. A place where this formalism will have immediate impact in the studies of hyperon-nucleon and hyperon-hyperon systems.


\noindent
\subsection*{Acknowledgments\footnote{RB dedicates this work to L\'azara Rufina Mart\'in Hern\'andez. Without the support she provided through the years this work would not have been possible. \\}  
 }
\noindent

RB acknowledges support from the U.S. Department of Energy contract DE-AC05-06OR23177, under which Jefferson Science Associates, LLC, manages and operates the Jefferson Laboratory.  RB would like to thank Martin Savage, Zohreh Davoudi, Thomas Luu, Robert Edwards, Kostas Orginos, Adam Szczepaniak, Jozef Dudek, Andr\'e Walker-Loud, Igor Danilkin, Maxwell Hansen, William Detmold, and Colin Morningstar for many useful discussions and feedback on previous versions of this manuscript. 

\bibliography{bibi}

 

\end{document}